\documentclass{article}
\usepackage[T1]{fontenc}
\usepackage{setspace}
\makeatletter

\newcommand{\LyX}{L\kern-.1667em\lower.25em\hbox{Y}\kern-.125emX\@}
\newenvironment{LyXParagraphIndent}[1]%
{
  \begin{list}{}{%
    \setlength\topsep{0pt}%
    \addtolength{\leftmargin}{#1}
    \setlength\parsep{0pt plus 1pt}%
  }
  \item[]
}
{\end{list}}

\usepackage[T1]{fontenc}
\font\bm=msbm10

\makeatother

\begin{document}

\title{A unified formal approach to weak and strong coupling expansions.}

\author{Beno\^it Bellet}

\maketitle
\begin{abstract}
The main issue of this work consists in extracting one or several
finite values for the sum of series involved in perturbation
theories. It is supposed to work for all cases in which two
physical parameters are involved, and makes thorough use of
dimensional arguments concerning these physical quantities. Weak
and strong coupling expansions are considered as the two faces of
a formal expression which is the central object of this method.
This so-called \emph{extension} procedure is systematic. We apply
it here to the divergent perturbative expansion of the ground
state energy of the anharmonic oscillator of quantum mechanics in
zero and one dimension, and provide, given a \(p\)-order divergent
expansion, an analytical expression of its estimated sum. This
method which is inspired by variational procedures provides high
degree of accuracy from lower orders of perturbation and seems to
have remarkable converge properties in a wide class of expansions
concerning physical observables.
\end{abstract}
\pagebreak

\section*{I. Introduction }

The problem of divergent series has long time been a big concern,
especially for physicists dealing with perturbative methods. In
many cases, the divergence of a perturbative expansion arises when
the solution of the free part of an equation is too "remote" from
the exact solution. The choice of this free part is theoretically
quite open but is actually restrained by the computability of the
free solution, together with its perturbative corrections.

When the series converges, the accuracy on the result only depends
on our ability to compute high order perturbation corrections.
Moreover, the convergence of the series and the fact that the
successive approximations tend to the right result, might be
guaranteed by mathematical theorems.

When the series diverges, the problem faced by the physicists is
one of dealing with an apparently worthless result (the truncated
series), despite the fact that physical arguments might ensure us
that this series actually corresponds to a known finite number.
The scientific literature is rich of various methods to treat this
situation: padé approximants \([1]\), variational methods
\({[2,3,4]}\), delta expansion \([5,6]\) with principle of minimal
sensitivity \([7]\) or multiple scale analysis \([8]\).

When facing the situation, one can note the following points:

\begin{enumerate}

\item While dealing with a physical problem, quantities with physical dimensions are
involved, and as a perturbative series expands in powers of a
dimensionless parameter, there is a first constraint on the way
physical quantities enter this dimensionless parameter. There is
also a second constraint regarding the dimension of the physical
quantity expanded itself.

\item When approximating a physical quantity in powers of the coupling constant, we
sometimes know a strong coupling expansion (in negative powers of
the coupling constant) to exist, even if it is not computable.

\end{enumerate}

These two points are central in the way I propose to deal with a
class of physical problems involving two physical parameters. The
existence of only two physical parameters leads to a unique way to
build up a dimensionless parameter \( g \) (say, the coupling
constant).

In the cases where the expression of a physical quantity in term
of the physical parameters is not exactly solvable, we expect two
different expansions of this physical quantity in powers of the
coupling constant to exist, at least formally: the weak coupling
expansion (in positive powers of \( g \)) and the strong coupling
expansion (in negative powers of \( g \)). Standard perturbative
method can give us access to the weak coupling expansion up to a
certain order \( p \). The case we are interested in in this paper
is the one when this expansion is divergent. The necessary
background for our original systematic method to work can now be
stated the following way:\\ the existence of a weak coupling
expansion of \( E \),

\begin{equation}
\label{wc} E( g )=\, \sum ^{\infty }_{N=0}E_{N}\, g^{N},
\end{equation}
\\and of a strong coupling expansion of \( E \),

\begin{equation}
\label{sc} \widetilde{E}( g )=\, g^{- \alpha/\delta}\, \sum
^{\infty }_{N=0}\widetilde{E}_{N}\, g^{- \gamma N/\delta},
\end{equation}
\\where \( E \) is a dimensionless function of the searched
physical quantity. The parameters \( \gamma \), \( \alpha \) and
\( \delta \) are crucial ingredients in the correct writing of the
srong coupling expansion. Their values usually be obtained
analysing the equation obeyed by the \( E \) (see Simon \([9]\)
for such a study on the energy of the anharmonic oscillator of
quantum mechanics). For our method to work on the calculation of
\( E \) , we thus need to know the following values:

\( \bullet  \)\,\, the \( E_{N} \) coefficients up to a certain
order \( p  \).

\( \bullet \)\,\, \( \gamma \), \( \alpha \) and \( \delta \)
 designed to be the smallest possible triplet of integers to satisfying eq.(\ref{sc}).
In the following, this triplet of integer values will be called
the \emph{signature} of \( E(g) \).\\

A good deal of literature has been devoted to variationnal methods
using principle of minimal sensitivity, which happen to be quite
efficient \({[3]}\). Some other attempts to deal with divergent
series through variationnal methods, making clever use of the
knowledge of the existence of a strong coupling expansion, were
successful \({[10]}\). The new method we introduce is original in
the way it considers weak and strong coupling expansions as two
faces of a single object called an \emph{extension}. Making use of
the formal properties of \emph{extensions}, one can obtain a
non-perurbative analytical estimate of \( E \) in term of the
coupling constant \( g \), given its signature and a certain
amount of coefficients of its divergent expansion. Note that the
method will also work as an accelerator of convergence for
convergent series.\\

The next section will introduce the extension method for the
trivial case where the signature is \emph{uniform}
(\(\gamma=\alpha=\delta=1\)) so that extensions write as
expansions in all positive and negative integer powers of the
coupling constant.
\\The completion of the general method through the
actual introduction of the notion of signature is performed in the
third section.
\\The fourth section is devoted to the detailed application of
the method to two specific cases. It is first experienced on the
divergent expansion of the zero dimension anharmonic oscillator.
The mathematical homogeneity of the integrated exponential
function allows us to consider the integral itself as depending on
two parameters with physical dimensions. The extension method will
happen to reproduce the exact result from the first order of
perturbation. As e second example, the method is applied to the
one dimensional anharmonic oscillator of quantum mechanics in the
calculation of the ground state energy. It will show a very
remarkable agreement with the exact result for any value of the
coupling constant.

\section*{II. Extensions of power series for uniform signatures}

In the simple case where the signature is uniform
(\(\gamma=\alpha=\delta=1\)), the respective weak and strong
coupling expansions of \(E\) in powers of a coupling constant
\(g\) write:
\[
 E( g )=\, \sum ^{\infty }_{N=0}E_{N}\, g^{N} \,\,\,\,\textrm{and}
 \,\,\,\, E( g )=g^{-1}\sum ^{\infty }_{N=0}\widetilde{E}_{N}\,
g^{-N}.
\]

We will consider these two expansion as parts of a single
\emph{extension}. To do so introduce such formal sums as:
\begin{eqnarray*}
Z_{0} & = & \sum _{n=-\infty }^{\infty }Z^{n}\\
Z_{1} & = & \sum _{n=-\infty }^{\infty }n\, Z^{n}\\
\\
\cdots  & = & \cdots \\
\\
Z_{i} & = & \sum _{n=-\infty }^{\infty }\frac{n!}{i!(n-i)!}Z^{n}\, \, \, .
\end{eqnarray*}
The manipulation of negative factorials is made algebraically
coherent introducing \((-1)!=\Omega\). For all coefficients for
which \(\Omega\) do not simplify, one let it go to infinity as
detailed at the beginning of the third part. When \( Z \) is a
given complex value out of the unit circle, these formal double
infinite series have the property to be convergent (resp.
divergent) in the positive powers of \( Z \) (called its
\emph{right part}) and divergent (resp. convergent) in the
negative powers of \( Z \) (called its \emph{left part}),
depending on the modulus of \(Z\).\\
\\
Note that \( Z_{i} \) can be expressed in term of \( Z_{0} \)
using derivatives:
\begin{equation}
\label{der}
Z_{i}=\frac{1}{i!}Z^{i}(\frac{d}{dZ})^{i}\, Z_{0}\, \, \, .
\end{equation}
\\
 They also have the following
formal property,
\[
i\geq 0,\, \, \, (1-Z)^{i+1}\, Z_{i}\equiv 0\, \, \, ,\]
 from
which one can logically expect the left expansion's sum (\( l \))
to be in some sense "opposite" to the right one (\( r \)). The
latter expectation can be easily checked as long as one performs
formal manipulations on power series without any consideration
about their convergence: Let us define the linear applications \(
r \) and \( l \).
\begin{eqnarray*}
r(Z_{i}) & = & \sum _{n=0}^{\infty
}\,\,\frac{n!}{i!(n-i)!}\,\,Z^{n}\\ l(Z_{i}) & = & \sum
_{n=-1}^{-\infty }\frac{n!}{i!(n-i)!}\,\,Z^{n}.
\end{eqnarray*}
The expression \( Z^{i}/(1-Z)^{i+1} \), when formally expanded in
powers of \(Z\), results in \(r(Z_{i})\), and when expanded in
powers  of \(1/Z\), results in \(-\,l(Z_{i})\). We write:
\begin{eqnarray*}
r(Z_{i}) & \sim &\, \, \, \, \, \, \frac{Z^{i}}{(1-Z)^{i+1}}\\
l(Z_{i}) & \sim &\, \, \, -\frac{Z^{i}}{(1-Z)^{i+1}}\, \, \, ,
\end{eqnarray*}
making sense of our previous remark about \(r(Z_i)\) and \(l(Z_i)
\) being 'opposite' to each other : \(r(Z_i)\sim -\,\,l(Z_i)\).\\
\\
In the following, we will combine elements as \( Z_{i} \)'s to
build a vector space. We hope any polynomial to have ``natural
extensions'' in this vector space.
\\ \\
\textbf{Definitions:}\\
\( \bullet  \)\,\,Let \( \Xi ^{p} \) be
the C-vector space of all polynomials with a lower than \( p \)
degree: \\
\[
A\in \Xi ^{p}\, \, \Longrightarrow A=\sum _{n=0}^{p}A_{n}\, Z^{n}\, \, \textrm{for}\, A_{n}\in \hbox {\bm \char'103 }\]
\\
\( \bullet  \)\,\,Let \( \Xi _{p} \) be the C-vector space
generated by extensions \( Z_{i} \) for \( i \) positive and lower
than \( p \):\\
\[
V\in \Xi _{p}\, \, \Longrightarrow V=\sum _{i=0}^{p}V^{i}\, Z_{i}\, \, \textrm{for }\, V^{i}\in \hbox {\bm \char'103 }
\]
\\
\subsection*{II.1. Extensions of a polynomial}

An element of \(\Xi_{p}\) will be considered as an extension of a
\(p\)-order polynomial if its restriction to zero to \(p\) powers
of \(Z\) identify to the given polynomial. We show now that this
properly defined extension of a polynomial is unique.

\textbf{Definition:}
\\ Let \( A \) and \( V \) be respective
vectors of \( \Xi ^{p} \) and \( \Xi _{p} \),
 then \( V \), once expanded as:
\[
V=\sum ^{\infty }_{n=-\infty }V_{n}\, Z^{n}\, \, \, ,
\]
is said to be \emph{a \( p \)-order extension of} \( A \) if:
\[
V_{n}=A_{n}\, \, \,\,\, \textrm{for}\,\,\, \, 0\leq n\leq p\, \,
\, .
\]
Moreover, if \(V=\sum ^{p}_{i=0}V^{i}\, Z_{i}\), from the
definition of the \( Z_{i} \)'s, we have:\\
\[
 V_{n}=\sum ^{p}_{i=0}\frac{n!}{i!(n-i)!}V^{i}\, \, \, .
\]\\

We will now proceed to show that such a \(p\)-extension of a given
polynomial is unique through the introduction of a specific linear
mapping between  \( \Xi ^{p} \) and \( \Xi _{p} \).
\\
\\
\textbf{Theorem 1:}\\ For given \( p \), let \( \varphi  \) and \(
\varphi ^{*} \) be the following linear mappings:
\[
\begin{array}{rrcl}
\varphi : & \Xi ^{p} & \longrightarrow  & \Xi _{p}\\
\,  &  &  & \\
 & Z^{n} & \longmapsto  & \sum ^{p}_{i=n}\frac{i!}{n!(i-n)!}(-1)^{n-i}Z_{i}\\
\,  &  &  & \\
\varphi ^{*}: & \Xi _{p} & \longrightarrow  & \Xi ^{p}\\
\,  &  &  & \\
 & Z_{i} & \longmapsto  & \sum ^{p}_{n=i}\frac{n!}{i!(n-i)!}Z^{n}\, \, \, .
\end{array}
\]
Then \( \varphi  \) and \( \varphi ^{*} \) are inverse linear
mappings\textbf{.}\\ \\ \textbf{Proof:}
\[
\begin{array}{rcl}
\varphi ^{*}(\varphi (Z^{n})) & = & \sum ^{p}_{i=n}\frac{i!}{n!(i-n)!}(-1)^{n-i}\varphi ^{*}(Z_{i})\\
\,  &  & \\
 & = & \sum ^{p}_{i=n}\, \sum ^{p}_{k=i}\frac{k!}{n!(i-n)!(k-i)!}(-1)^{n-i}Z^{k}
\end{array}
\]
Using the following identity for \( k\leq p \), where \( \delta  \) is the
Kroeneker symbol:
\[
\sum ^{p}_{i=n}\frac{1}{(i-n)!(k-i)!}(-1)^{n-i}=\delta _{n,k}\, \,
\, ,
\]
we prove that for \( 0\leq n\leq p \) and \( 0\leq i\leq p\):
\[
\varphi ^{*}(\varphi (Z^{n}))=Z^{n}\, \, \, \, \, \, \,
\textrm{and}\, \, \, \, \, \, \, \varphi (\varphi
^{*}(Z_{i}))=Z_{i}\, \, \, .
\]

>From the existence of the reversible mapping \( \varphi  \)
between \( \Xi ^{p} \) and \( \Xi _{p} \) and the fact that \(
\{Z^{n}\}_{0\leq n\leq p} \) is a base of \( \Xi ^{p} \), it comes
that \( \{Z_{i}\}_{0\leq i\leq p} \) is a base of \( \Xi _{p}
\).\\
\\
\textbf{Theorem 2:}
\\
Let \( A\in \Xi ^{p} \), then \( \varphi (A)\in \Xi _{p} \) is the
only extension of \( A \).\\
\\
\textbf{Proof:}
\\
Let us first recall that:
\[
Z_{i}=\sum _{n=-\infty }^{\infty }\frac{n!}{i!(n-i)!}Z^{n}
\]
\[
\varphi ^{*}(Z_{i})=\sum _{n=0}^{p}\frac{n!}{i!(n-i)!}Z^{n}\, \, \, ,
\]
\\
such that by construction, for \( 0\leq i\leq p \), \( Z_{i} \) is
a \( p \)-order extension of \( \varphi ^{*}(Z_{i}) \). \\
\\
If now \( A=\sum ^{p}_{n=0}\, A_{n}\, Z^{n} \), one can write \( \varphi (A)=\sum ^{p}_{i=0}\, A^{i}\, Z_{i} \)
where:
\begin{equation}
\label{tra}
A^{i}=\sum ^{i}_{n=0}\frac{i!}{n!(i-n)!}(-1)^{n-i}A_{n}\, \, \, .
\end{equation}
\\
>From the reversibility of \( \varphi , \) one has \( \varphi (\varphi ^{*}(A))=A \),
and we can establish that:
\[
\begin{array}{rcl}
A & = & \sum ^{p}_{i=0}A^{i}\, \varphi ^{*}(Z_{i})\\
\,  &  & \\
\varphi (A) & = & \sum ^{p}_{i=0}A^{i}\, Z_{i}\, \, \, .
\end{array}
\]
As \( Z_{i} \) is an extension of \( \varphi ^{*}(Z_{i}) \), it follows that
\( \varphi (A) \) is an extension of \( A \) for any \( A\in \Xi ^{p} \).
The reversibility of \( \varphi  \) guarantees the unicity of this extension.
\\
\\
\textbf{Example:}
\\
For \( p=2 \), let us find the extension of the \( 2 \)-degree
polynomial:
\[
A=1+2Z+3Z^{2}\, \, \, .\]
The coordinates of its second order
extension are given by the relation (\ref{tra}):
\[
\begin{array}{rclcl}
A^{0} & = & A_{0} & = & 1\\
A^{1} & = & -A_{0}+A_{1} & = & 1\\
A^{2} & = & A_{0}-2A_{1}+A_{2} & = & 0\, \, \, ,
\end{array}\]
so that:
\\
\( \bullet  \)\,\, \( \varphi (A)=A^{0}\, Z_{0}+A^{1}\, Z_{1} \)
is the extension of \( A \); it also writes:
\[
\varphi (A)=\sum ^{\infty }_{n=-\infty }(n+1)\, Z^{n}\, \, \, ,
\]
and is actually a second order extension of \( A \), as expected.
\\
\( \bullet  \)\,\, \( r(\varphi
(A))=r(Z_{0})+r(Z_{1})\sim1/(1-Z)+Z/(1-Z)^{2}=1/(1-Z)^{2} \) is
what one could reasonably expect as a sum for a divergent series
behaving like \( A \) does for lower orders.
\\
\\
\textbf{Remark:}
\\
This example illustrates an important case where the extension of
a \(p\)-order polynomial (in \(\Xi^p\)) actually belongs to
\(\Xi^{p-1}\), that is \(A^p=0\)
\\

The latter remark will be of great use in the following introduction of a parameter
in our method. It will also be verified as a test for the reliability of the
method in the example of application given in the fourth section.

\subsection*{II.2. The introduction of a parameter: \protect\( \omega \protect \)}

We will show that the simple introduction of a complex parameter
\( \omega  \) allows us to build as many \( \Xi _{p} \) vector
spaces which means as many \(p\)-order extensions of a given
polynomial as non vanishing values for \( \omega  \). There is
then no reason why restraining to the previous \( \omega =1 \)
case. At the end, the problem will just remain to find the most
appropriate value of \( \omega  \). \\
 \\
>From now, for a fixed \( \omega  \), \( \Xi ^{p} \) and \( \Xi _{p} \) will
be redefined:
\[
A\in \Xi ^{p}\, \, \Longrightarrow A=\sum _{n=0}^{p}A_{n}\, (\omega Z)^{n}\, \, \textrm{for}\, \omega \in \hbox {\bm \char'103 }^{*}\, \textrm{and}\, A_{n}\in \hbox {\bm \char'103 }\]
\[
V\in \Xi _{p}\, \, \Longrightarrow V=\sum _{i=0}^{p}V^{i}\, (\omega Z)_{i}\, \, \textrm{for }\, \omega \in \hbox {\bm \char'103 }^{*}\, \textrm{and}\, V^{i}\in \hbox {\bm \char'103 }\]
where \( (\omega Z)^{n}=\omega ^{n}\, Z^{n} \), and \( (\omega Z)_{i} \) is
defined as follows, \\
\[
\begin{array}{rcl}
(\omega Z)_{i} & = & \sum _{n=-\infty }^{\infty }\frac{n!}{i!(n-i)!}(\omega Z)^{n}
\end{array}\]
\( \varphi  \) now acts on vectors as \( (\omega Z)^{n} \):
\[
\begin{array}{rrcl}
\varphi : & \Xi ^{p} & \longrightarrow  & \Xi _{p}\\
\,  &  &  & \\
 & (\omega Z)^{n} & \longmapsto  & \sum ^{p}_{i=n}\frac{i!}{n!(i-n)!}(-1)^{n-i}\, (\omega Z)_{i}
\end{array}\]
\\
Theorems 1 and 2 still hold as long as \( \omega  \) does not vanish and the
method now produces an extension \( \varphi (A) \) of \( A \), depending on
the value of \( \omega  \):
\[
A=\sum ^{p}_{n=0}A_{n}\, Z^{n}=\sum ^{p}_{n=0}\frac{A_{n}}{\omega ^{n}}\, (\omega Z)^{n}\, \, \, ,\]
\[
\varphi (A)=\sum ^{p}_{n=0}\frac{A_{n}}{\omega ^{n}}\, \varphi ((\omega Z)^{n})=\sum ^{p}_{i=0}A^{i}(\omega )\, (\omega Z)_{i}\, \, \, ,\]
where
\[
A^{i}(\omega )=\sum ^{i}_{n=0}\frac{i!}{n!(i-n)!}(-1)^{n-i}\frac{A_{n}}{\omega ^{n}}\, \, \, .\]
\\

When trying to extend a given \( p \)-degree polynomial \( A \),
one now has to deal with as many extensions as possible values for
\( \omega  \). The last remark of the preceding section suggests
that we could use this freedom on the value of the parameter \(
\omega  \) to make \( A^{p}(\omega ) \) vanish so that we could
just keep \(p\)-order extensions of \( A \) that lie in \( \Xi
_{p-1} \). \( A^{p}(\omega ) \) being a \( p \)-degree polynomial
in \( \omega \), one expects \( \omega  \) to take at most \( p \)
values (\( A^{p}(\omega )=0 \)) so that a vector of \( \Xi ^{p} \)
will have at most \( p \) different extensions. If \( \varpi  \)
is a root of \( A^{p}(\omega ) \), it solves:
\begin{equation}
\label{opt}
A^{p}(\varpi )=\sum ^{p}_{n=0}\frac{p!}{n!(p-n)!}(-1)^{n-p}\, \frac{A_{n}}{\varpi ^{n}}=0\, \, \, ,
\end{equation}
 and the associated extension writes:\\
\begin{equation}
\label{ext}
\varphi (A)\, =\, \sum ^{p-1}_{i=0}\, A^{i}(\varpi )\, (\varpi Z)_{i}\, \, \, .
\end{equation}
 \\
The problem now remains to choose from all these extensions.
\\
\subsection*{II.3. A criterion for the choice: \protect\( \omega =\varpi \protect \)}

The aim of this work is, given a divergent series up to order \( p
\), to obtain an estimation of its complete sum. Let us start with
the method applied to the \( (p-1) \)-order. In order to establish
a convenient criterion for the choice of the ``better'' \( \omega
\) between the roots of \( A^{p-1}(\omega ) \), we will look for
the one which makes the better prediction for the forthcoming term
of the series: \( A_{p} \). \\ To do so, let us suppose that \(
A^{p-1}(\omega ) \) and \( A^{p}(\omega ) \) have a common root \(
\varpi  \), the following properties come:
\\
\( \bullet  \)\,\, From the equation (\ref{ext}), the \( \varpi
\)-associated \( (p-1) \)-order and \( p \)-order extensions
identify. As a consequence, the forthcoming term predicted by the
\( (p-1) \)-order extension is the exact value \( A_{p} \).
\\
\( \bullet  \)\,\, From the definition (\ref{opt}) of \(
A^{p}(\omega ) \), one can establish the following relation:
\[
\frac{d}{d\omega }(\omega ^{p}\, A^{p}(\omega ))=-p\, \omega
^{p-1}\, A^{p-1}(\omega )\, \, \, ,
\]
so that if \( \varpi  \) is
a common root of \( A^{p-1}(\omega ) \) and \( A^{p}(\omega ) \),
it has to be a double root of \( A^{p}(\omega ) \).
\\

The goal is to achieve the best estimate of the complete sum of
the series through its extension at a finite order. It seems
natural to be interested in the case where two successive
extensions (orders \(p-1\) and \(p\)) are equal, thus at order
\(p\), double roots are our best candidate for the choice of
\(\varpi\). In the absence of double roots, we decide to look for
the closest couple of roots - in terms of distance in the complex
plane - and select one of these. In doing so, we expect our
selected root to be also close to a root of the previous estimate.
As a consequence, these two successive extensions are hoped to be
close to each other. The applications of the method in section IV
will corroborate, through numerical observations, these general
remarks.

\section*{III. Signature of series in physics}

For the moment, to keep reasonable, one has to expect most of the
series encountered in physics not to fit in this frame in the
sense that the predictions for the forthcoming terms of a
development are far from the right one (and that the roots are far
from each other). Let us consider for example such a ``misfit''
expansion:
\[
\sum ^{p}_{n=0}\frac{1}{(n/2)!}\, Z^{n}\, \, \, ,
\]
which expected large order behaviour can obviously not be found as
a combination of the \(Z_{i}\)'s as defined. The fact that
forthcoming terms of such a series will be very badly predicted by
our method can also be pointed out when noticing that an extension
of this truncated series has to vanish for even and negative
values of \( n \) (as factorials of negative integers diverge).
The \(Z_{i}\)'s as defined can obviously not satisfy such a
condition. This section will take into account this kind of
difficulty (now to be considered as a clue) when introducing the
notion of signature relying on the knowledge we have of the form
of the strong coupling expansion.

Given a finite series of complex numbers, we now have at our
disposal a method which permits to extend it infinitely together
in positive and negative powers of the coupling constant \( Z \).
Moreover, the right and left parts of this extension can be summed
formally. Some of the series we are interested in are
characterized by cancellations of an infinite set of coefficients
(e.g. \( A_{n}=0 \) for \( n \) even and negative. We shall later
concentrate on how the occurrence of the vanishing coefficients is
in related to the form of the strong coupling expansion. Since no
combination of the \(Z_i\)'s can reproduce this kind of behaviour,
we ask ourselves now what to do then when we know the expected
series to have such a typical property? In this paper, we shall
consider such cases where the occurrence of vanishing terms is
regular. Let us first consider the following double infinite
series where by construction all even and negative powers of the
coupling constant do vanish:
\begin{equation}
\label{T12}
T=\sum ^{\infty }_{n=-\infty }T_{n}\, Z^{n}\, \, \, \, \, \, \textrm{where}\, \, \, \, \, T_{n}=\frac{(-1-n)!((n-1)/2)!}{(-1)!}
\end{equation}
\\
As previously mentioned, it will be of great use to reconsider the
definitions of the factorials in order to make sense with
expressions as \( (-2)!/(-1)!=-1 \):

\begin{LyXParagraphIndent}{1cm}
~\\
Definition:
\\
\( \bullet  \) \( (n-1)!\, n=n! \) ~~~~~for ~~~~~\( n\in \hbox
{\bm \char'121 }^{*} \),\\ \( \bullet  \) \( (-1)!=\Omega  \)
~~~~~is infinitely large,\\ \( \bullet  \) \( n!=\Gamma (n+1) \)
~~~~~for~~~~\( n\in [0,1]\cap \hbox {\bm \char'121 } \).
 ~\\
\\
The following properties come:
\\
\( \bullet  \) \( n!\, (-1-n)!\, \sin (n\pi )=\pi  \)~~~~~for~~~~~~\( n\in \hbox {\bm \char'121 } \)
~~~~and~~~~~\( n\notin \hbox {\bm \char'132 } \),\\
\( \bullet  \) \( n!\, (-1-n)!\, (-1)^{n}=\Omega  \) ~~~~~for~~~~~\( n\in \hbox {\bm \char'116 } \).

\end{LyXParagraphIndent}
~\\ Using these properties, the only non-vanishing terms of \( T
\) are:
\\
\[
T_{N}=\frac{((N-1)/2)!}{N!}(-1)^{N}\, \, \, \, \, \textrm{and}\, \, \, \, \, T_{-1-2N}=\frac{(2N)!}{N!}(-1)^{N}\, \, \, \, \, \textrm{for}\, \, \, \, \, N\in \hbox {\bm \char'116 }
\]
 \\
When looking at the left part of \(T\), these cancellations are
typical of what has been called the signature of \(T\), as it
points out all the vanishing terms of the expansion it corresponds
to.

~\\ \textbf{Definitions:}
\\
\( \bullet  \)\,\, If \( T \) is such that its non-vanishing terms
are \( T_{\delta N} \) and \( T_{-\alpha -\gamma N} \) for \( N\in
\hbox {\bm \char'116 } \), then it will be said to have a \(
\gamma |\alpha |\delta  \)\emph{-signature}. A typical example of
this signature is provided by:
\[
T_{n}=\frac{(-1-\frac{n}{\delta })!(-1+\frac{\alpha +n}{\gamma
})!}{\Omega }\, \, \, .
\]
For a given signature, this \(T\) will be called its associated
\emph{transformation}.
\\
As a matter of example, the above series (\ref{T12}) has a \(
2|1|1 \)-signature and the \( 1|1|1 \)-signature considered in the
preceding section was simply called uniform. \\
\\
\( \bullet  \)\,\, Let us now define the multiplication \( T*A \):
\[
A=\sum ^{\infty }_{n=-\infty }A_{n}\, Z^{n}\, \, \,  \Rightarrow
\, \, \, T*A=\sum ^{\infty }_{n=-\infty }T_{n}\, A_{n}\, Z^{n}
\]
 then, if \( A \) has a uniform signature, \( T*A \) will have a \( \gamma |\alpha |\delta  \)\emph{-}signature
through the action of the transformation \( T \).\\
\\
\( \bullet  \) If \( B \) is an extension of \( A \), then \( T*B
\) is a \emph{\( T \)-extension} of \( T*A \).

\subsection*{III.1. The case of non-uniform signatures}

As it is formatted to give rise to extensions with uniform
signatures, the method developed in the preceding section
obviously needs a few adjustment to fit the cases where the
signature is known not to be uniform. Suppose we wish to extend a
polynomial \( B \) known to be associated to a non-uniform
signature. The method now consists in ``removing'' its signature
multiplying it by \( T^{-1} \), and let it be extended (through
the action of \( \varphi  \)) into an extension with uniform
signature. Then, we just have to put back the proper signature
multiplying the result by the associated transformation \( T \).
We thus make sure that the \( T \)-extension obtained has the
expected signature (respecting the expected vanishing terms), and
is actually a \( T \)-extension of the original polynomial.\\
\\
If \( B \) is associated to a \( \gamma |\alpha |\delta
\)-signature, then it expands in powers of \( Z^{\delta } \) in
the weak coupling limit and writes up to order \( p \):
\[
B=\sum ^{p}_{n=0}B_{\delta n}\, Z^{\delta n}
\]
Where the connection with equation (\ref{wc}) is obtained
replacing \(g\) with \(Z^{\delta}\). \( B \) can then be rewritten
\( T*(T^{-1}*B) \) where \( T^{-1}*B \) can be associated to a
uniform signature:
\[
T^{-1}*B=\sum ^{p}_{n=0}T_{\delta n}^{-1}\, B_{\delta n}\, Z^{\delta n}\]
The extension of \( T^{-1}*B \) can now be found using the method of the second
section. \( \varphi (T^{-1}*B) \), its extension, will have a uniform signature.
The latter product with \( T \) ensures us to recover the proper \( \gamma |\alpha |\delta  \)-signature
for the \( T \)-extension of \( B \):
\[
T*\varphi (T^{-1}*B)\, \, \, \, \, \textrm{is a }T-\textrm{extension of}\, \, \, \, \, B
\]
 \\
\textbf{Definition:}
\\
In order to achieve the previous manipulation \( \varphi  \) and
\( \varphi ^{*} \) have to be formally updated in replacing \( Z
\) by \( Z^{\delta } \) in their definitions so that they remain
inverse mappings. The formal expressions then arising are made
explicit as follows:
\[
(Z^{\delta })^{n}=Z^{\delta n}\, \, \, \textrm{and}\, \, \,
(Z^{\delta })_{i}=\sum ^{\infty }_{n=-\infty }\frac{(n/\delta
)!}{i!\,(n/\delta -i)!}\, Z^{n}
\]
Using this procedure, we can now obtain the unique extension of
any polynomial given in addition to its coefficients, its
non-uniform signature.
\\

Our goal is at the end to obtain a non-perturbative expression
corresponding to the physical quantity corresponding to \(B\) up
to a finite order. To do so, we just have to calculate the sum
whether of the right part or of the left part of an extension with
a non-uniform signature. Both these sums will reveal to equate up
to a simple multiplicative term. If one of them is convergent, the
other is not and we will have to consider this second one as
formal.
\\

\subsection*{III.2. Summation }

In the general case of a \( \gamma |\alpha |\delta  \)-signature, left and
right parts of a \( T \)-extension \( \overline{B} \) of \( B \) write:
\[
r(\overline{B})=\sum ^{\infty }_{N=0}\overline{B}_{\delta N}\,
Z^{\delta N}\, \, \, \textrm{and}\, \, \, l(\overline{B})=\sum
^{\infty }_{N=0}\overline{B}_{-\alpha -\gamma N}\, Z^{-\alpha
-\gamma N}
\]
The \( p \)-order \( T \)-extension \( \overline{B} \) writes \(
T*A \) where \( T \) is the transformation associated to its
signature, and \( A=\varphi (T^{-1}*B) \) is a linear combination
of vectors \( ((\omega Z)^{\delta })_{i} \):\\
\[
T_{n}=\frac{(-1-\frac{n}{\delta })!(-1+\frac{\alpha +n}{\gamma })!}{\Omega }
\]

\[
A=\sum ^{p-1}_{i=0}A^{i}(\omega )\, ((\omega Z)^{\delta })_{i}\,
\, \, \textrm{where}\, \, \, \omega \, \textrm{satisfies}\,
A^{p}(\omega )=0
\]
 \\
Now recall from equation (\ref{der}) a useful relation between
extensions:
\begin{equation}
((\omega Z)^{\delta })_{i}=\frac{1}{i!}Z^{\delta i}(\frac{d}{dZ^{\delta }})^{i}\, ((\omega Z)^{\delta })_{0}
\end{equation}
This property suggests to rewrite \( T*A \) as:
\begin{equation}
\label{T*A}
 T*A=\sum ^{p-1}_{i=0}A^{i}(\omega
)\frac{1}{i!}Z^{\delta i}(\frac{d}{dZ^{\delta }})^{i}\, \left[
T*((\omega Z)^{\delta })_{0}\right]
\end{equation}
Then, from the sums of \( T*((\omega Z)^{\delta })_{0} \), the
sums of \( T*A \) will follow for any extension \( A \). But as \(
(Z^{\delta })_{0}=Z_{0} \) and \( T*Z_{0}=T \), we simply have to
sum the left and right parts of \( T \).
\\
\\
\textbf{The summation of \( T: \)}
\\
\[
\begin{array}{rclrcl}
l(T) & = & \sum ^{\infty }_{N=0}T_{-\alpha -\gamma N}\, Z^{-\alpha -\gamma N} &  & = & \sum ^{\infty }_{N=0}(-1)^{N}\frac{(-1+(\alpha +\gamma N)/\delta )!}{N!}\, Z^{-\alpha -\gamma N}\\
 & \,  &  &  & \,  & \\
r(T) & = & \sum ^{\infty }_{N=0}T_{\delta N}\, Z^{\delta N} &  & = & \sum ^{\infty }_{N=0}(-1)^{N}\frac{(-1+(\alpha +\delta N)/\gamma )!}{N!}Z^{\delta N}
\end{array}
\]
\\
Let us now introduce integral expressions for fractional
factorials where \( N \) and \( \nu  \) are respectively positive
and non zero positive integers:
\[
(N/\nu )!=\nu \: \int ^{\infty }_{0}dt\, t^{\nu -1}\, e^{-t^{\nu
}}\, t^{N}\, \, \, .
\]
If we accept to permute the sum and the
integral without wondering whether they diverge or not, left and
right parts of \( T \) can be formally summed:
\[
l(T)\sim \delta \, \int ^{\infty }_{0}\frac{dt}{t}e^{-(t/Z)^{\gamma }-t^{\delta }}(t/Z)^{\alpha }\, \, \, \, \, \textrm{and}\, \, \, \, \, r(T)\sim \gamma \, \int ^{\infty }_{0}\frac{dt}{t}e^{-t^{\gamma }-(Zt)^{\delta }}t^{\alpha }
\]
The link between left and right parts of \( T \) comes from a simple change
of variable in the integral:
\begin{equation}
\label{l&r} \gamma \, l(T)\, \sim \, \delta \, r(T).
\end{equation}
The analytical properties of \( l \) and \( r \) as functions of
\( Z \) are the one of multi-valued functions. It has also to be
noted that, depending on the values of \( \gamma  \), \( \alpha \)
and \( \delta  \), the integral expression does not always make
sense as conditions on its existence might not be satisfied.
Nevertheless, as \( \gamma  \) and \( \delta  \) are positive
integers, only a strictly negative value for \( \alpha  \) can
make the integral diverge around \( t=0 \). If this situation
occurs, then the first terms of the expansion \( l(T) \) and \(
r(T) \) (in respective powers of \( Z^{-\gamma } \) and \(
Z^{\delta } \)) will not be incorporated in the integral
expression. As only a finite number of terms are concerned, this
will be a complication but not an obstacle to the summing of the
series. The application of the method given in the fourth section
will illustrate this situation.
\\
\\
\textbf{The summation of \,\,\( T*A \)}
\\
The summed expression for \( r(T*A) \) is now from (\ref{T*A}):
\begin{equation}
\label{sum}
\begin{array}{rcl}
r(T*A) & \sim  & \gamma \, \sum ^{p-1}_{i=0}A^{i}(\omega )\frac{1}{i!}Z^{\delta i}(\frac{d}{dZ^{\delta }})^{i}\, \int ^{\infty }_{0}\frac{dt}{t}e^{-t^{\gamma }-(\omega Zt)^{\delta }}t^{\alpha }\\
 & \,  & \\
 & \sim  & \gamma \, \int ^{\infty }_{0}\frac{dt}{t}e^{-(t/\omega Z)^{\gamma }-t^{\delta }}(t/\omega Z)^{\alpha }\, \sum ^{p-1}_{i=0}A^{i}(\omega )\frac{1}{i!}(-t^{\delta })^{i}
\end{array}
\end{equation}
where one has successively performed all the derivatives and then
a change of variable. The first expression has, by construction,
the correct expansion in powers of \( Z^{\delta } \) up to order
\( p \). The relation (\ref{l&r}) between left and right parts of
\( T*A \) still holds as \( l \) and \( r \) are linear mappings:
\[
\gamma \, l(T*A)\, \sim \, \delta \, r(T*A)
\]
The formal manipulation performed to obtain the second expression
of (\ref{sum}) is such that its expansions around \( Z\sim \infty
\) and \( Z=0 \) will respectively identify to \( \frac{\gamma
}{\delta }\, l(T*A) \) and \( r(T*A) \). This last point is
crucial in the way it it guarantees that the form of the left and
right parts (the strong and weak coupling expansions) will be
properly recovered.
\\
The appendix will concatenate the step by step expansion procedure
to get an extension and a sum from the knowledge of a \( p
\)-order series and its signature. The next section will
illustrate the method on the divergent perturbative series of
quantum mechanics in zero and one dimension.

\section*{IV. Applications of the extension method}

The extension method will now be applied to the case of the zero
dimensional anharmonic oscillator, which happens to be a simple
integral. It will then be tested on the calculation of the ground
state energy of the quantum mechanics anharmonic oscillator. These
applications will follow the extension procedure as it is
concatenated in the appendix (the signature, the extension and the
sum).

\subsection*{IV.1. The zero dimension anharmonic oscillator}

We consider the following integral:
\\
\[
E(m,\lambda)=\sqrt{2}\int^{\infty}_{0} dx \,\,e^{-m
x^2/2\,-\lambda x^4/4},
\]
which, when expanded in powers of \(\lambda/m^2\) gives rise to a
divergent series.
\\
\\
\textbf{The Signature:}
\\
A dimensional analysis on the base of the zero-degree homogeneity
of the exponential function gives:
\[
[E]\,\,=\,\,[x]\,\,=\,\,[m^{-1/2}]\,\,=\,\,[\lambda^{-1/4}],
\]
where \(m\) and \(\lambda\) are to be considered as the two
parameters with a physical dimension. Then \(E\) will expand both
in powers of \(\lambda/m^2\) in the weak coupling limit, and in
powers of \((\lambda/m^2)^{-1/4}\) in the strong coupling limit:
\\
\[
E(\lambda )=m^{-1/2}\, \sum ^{\infty }_{N=0}E_{N}\, (\frac{\lambda
}{m^{2}})^{N},
\]
\[
\widetilde{E}(\lambda )=\lambda ^{-1/4}\, \sum ^{\infty
}_{N=0}\widetilde{E}_{N}\, (\frac{m}{\lambda ^{1/2}})^{N}.
\]
Note that \(E(\lambda )\) is divergent and \(\widetilde{E}(\lambda
)\) is convergent.
\\
The signature in then given by (\( \gamma =2,\, \alpha =1,\,
\delta =4 \)). Taking \(\lambda/m^2=Z^4\) and \(m=1\), right and
left expansions of such an expression respectively rewrite as:
\[
E=\sum ^{\infty }_{N=0}E_{N}\, Z^{4N}\, \, \, \textrm{and}\, \, \,
\widetilde{E}=Z^{-1}\, \sum ^{\infty }_{N=0}\widetilde{E}_{N}\,
Z^{-2N}\,\,,
\]
exhibiting a \( 2|1|4 \)-signature.
\\

 Given the first coefficients
of the right expansion, let us apply the extension procedure to
the first order of this divergent series (now called \(B\)):
\[
B=B_{0}+B_{4} Z^{4} +\, . \, . \, .
\]
where
\[
B_{0}=\sqrt{2}\int^{\infty}_{0} dx \,\,e^{-m
x^2/2}=\sqrt{\pi}
\]
and
\[
B_{4}=-\sqrt{2}\int^{\infty}_{0} dx \,\,e^{-m x^2/2}\,\,
\frac{x^4}{4}=-\frac{3}{4}\sqrt{\pi}.
\]
We now need the transformation associated to the \( 2|1|4
\)-signature:
\[
T_{n}==\frac{(-1-\frac{n}{4 })!(-1+\frac{1+n}{2})!}{\Omega },
\]
such that:
\[
T_{0}=\sqrt{\pi}\,\,\,\textrm{and}\,\,\,T_{4}=-\frac{3}{4}\sqrt{\pi}.
\]
The first two coefficients of the uniform expansion (freed from
its signature) to be extended write:
\[
A_{0}=T^{-1}_{0}B_{0}=1\,\,\,\textrm{and}\,\,\,A_{4}=T^{-1}_{4}B_{4}=1.
\]
\\
\textbf{The Extension:}
\\
The parameter \(\omega\) has to satisfy the equation:
\[
A^{1}(\omega)=-1+\frac{1}{\omega^4}\,\,\,\Longleftrightarrow\,\,\,\omega^{4}=1
\]
We have now all the necessary material to give an estimation of
the whole sum, given its first two terms and its signature.
\\
\\
\textbf{The Sum:}
\\
The first order estimation of the integral through the extension
procedure is then:
\[
2\int^{\infty}_{0} dt \frac{1}{Z} \,\,e^{-(t/Z)^2\,-t^4},
\]
which also writes (taking \(t=\sqrt{m}\,Z\,x/\sqrt{2}\,\) and \(
Z=(\lambda/m^2)^{1/4} \)):
\[
\sqrt{2}\int^{\infty}_{0} dx \,\,e^{- m x^2/2\,-\lambda x^4/4}.
\]
This an exact result.

 As a matter of comparison: in the case of
the calculation of this integral, the usual variationnal method
with principal of minimal sensitivity happens to provide a
non-exact finite result from any of its \(p\)-order divergent
expansion \([2,4]\) while the extension method gives the exact
result from the first order of perturbation. It is now
straightforward to notice that any integral of the form:
\[
\int^{\infty}_{0} dx \,\,x^{\alpha-1}\,\,e^{-
x^{\gamma}\,-Z^{\delta} x^{\delta}}
\]
will also be exactly estimated form its first order of expansion
in powers of \(Z^{\delta}\) using the extension procedure. This
class of integral can thus be considered as canonical regarding
the extension method.

\subsection*{IV.2. The ground state energy of the anharmonic
oscillator} The development of the vacuum energy \( E \) of the
anharmonic oscillator of quantum mechanics is known to be
divergent when expanded in powers of the coupling constant \(
\lambda  \). Many papers tried with success to deal with this
divergent series, sometimes with high level of accuracy and/or
complexity \([3,7,10,11] \).
\\
The Lagrangian of the model writes:
\[
L(m^{2},\lambda )=\frac{1}{2}(\partial _{t}\Phi )^{2}-\frac{1}{2}m^{2}\Phi ^{2}-\frac{1}{4}\lambda \Phi ^{4}
\]
A dimensional analysis of physical quantities gives:
\[
[L]=1\, \, \, ;\, \, \, [t]=-1\, \, \, ;\, \, \, [\Phi ]=-1/2\, \, \, ;
\]
\[
[m]=1\, \, \, ;\, \, \, [\lambda ]=3\, \, \, ;\, \, \, [E]=1\, \,
\, .
\]
The vacuum energy for the free Lagrangian \( E(m^{2},\lambda
\simeq 0) \) and its perturbative corrections in powers of \(
\lambda  \) are computable, but one can also imagine another
perturbative expansion in powers of \( m^{2} \) around the free
Lagrangian \( L(m^{2}\simeq 0,\lambda ) \). This latter strong
coupling expansion, although proved to exist and be convergent \(
[9] \), is not computable with the usual technique of perturbation
as one cannot solve the free equation. Nevertheless, from
dimensional analysis, we know that these two expansions
respectively write:
\[
E(\lambda )=m\, \sum ^{\infty }_{N=0}E_{N}\, (\frac{\lambda }{m^{3}})^{N}
\]
for the divergent expansion around \( \lambda =0 \) and,
\[
\widetilde{E}(\lambda )=\lambda ^{1/3}\, \sum ^{\infty }_{N=0}\widetilde{E}_{N}\, (\frac{m^{2}}{\lambda ^{2/3}})^{N}\]
for the convergent expansion around \( m^{2}=0 \) (or \( \lambda \sim \infty  \)).\\
\\
Without loosing any dimensional information, one can get rid of fractional powers
writing \( \lambda /m^{3}=Z^{3} \) and \( m=1 \), then:
\[
E=\sum ^{\infty }_{N=0}E_{N}\, Z^{3N}\, \, \, \textrm{and}\, \, \, \widetilde{E}=Z\, \sum ^{\infty }_{N=0}\widetilde{E}_{N}\, Z^{-2N}
\]
Considering both these series as right (\( E \)) and left (\(
\widetilde{E} \)) parts of the same object we can recognize an
extension with \( 2|-1|3 \)-signature.
\\
We can now use the extension procedure to sum the known divergent
expansion of \( E \) to a finite order \( p \) on the grounds of a
\( 2|-1|3 \)-signature (\( \gamma =2,\, \alpha =-1,\, \delta =3
\)). The perturbative series of the ground state energy \( E \) to
be extended will be called \( B \). A systematic procedure
computes the coefficients of the ground state energy (\( B_{3n}
\))  \( [12] \):
\[
B=\frac{1}{2}+\frac{3}{16}Z^{3}-\frac{21}{128}Z^{6}+\frac{333}{1024}Z^{9}-\frac{30885}{32768}Z^{12}+\frac{916731}{262144}Z^{15}+.\, .\, .
\]
Left and right parts of its extensions are to be respective
approximations of \( \frac{3}{2}\widetilde{E} \) and \( E \).

\subsubsection*{IV.2.1. The choice of the ``best'' root}

The search for a \( p \)-order \( T \)-extension goes through the
choice between \( p \) roots of \( A^{p}(\omega ) \): \( \omega
_{1},\, \omega _{2},\, .\, .\, ,\omega _{p} \) . The discussion of
the section II.3 suggests to consider the root which is as close
as possible to another one of the same order (a similar comparison
with the preceding order roots reveals to be equivalent). In order
to measure this property of a given root \( \omega _{i} \) to be
close to one or several other roots of the same order, we
introduce its \emph{weight}:
\[
weight(\omega _{i})=\sum _{j\neq i}\frac{1}{(\omega _{j}-\omega _{i})^{2}}
\]
The following array will give the roots of \( A^{p}(\omega ) \)
for \( p=2,\, 3,\, 4,\, \textrm{and}\, 5 \) together with their
weights (in brackets):
{\small\[
\begin{array}{llll}
p & \omega _{i}\, (weight) &  & \\
2 & \underline{1.062}\, (191) & 1.135\, (191) & \\
3 & \underline{1.053\pm 0.061i}\, (113) & 1.189\, (90) & \\
4 & 1.043\pm 0.102i\, (135) & \underline{1.071}\, (216) & 1.236\, (79)\\
5 & 1.034\pm 0.14i\, (135) & \underline{1.073\pm 0.033i}\, (355) & 1.279\, (71)
\end{array}
\]}
For every order, the root associated to the biggest weight are
underlined (the choice for the underlined root of the second order
is arbitrary as both of them have the same weight). \\ As \(
A^{p}(\omega ) \) is in fact a polynomial in powers of \( 1/\omega
^{3} \), it has \( p \) triplets of roots.  is why we decided to
consider only one root in each triplet: the one lying around the
real axis. This choice will not affect the right part of the
extension.
\\

In order to test our selected roots, a similar array will show the
related estimations for the forthcoming term of the series (as the
\( p \)-order \( T \)-extension provides an estimation for the
known value of \( B_{3p+3} \)). The estimations associated to the
chosen roots will be underlined and the last column will show the
exact value: {\small\[
\begin{array}{lllll}
p & \overline{B}_{3p+3}\, (est.) &  &  & B_{3p+3}\, (exact)\\
2 & \underline{0.3223} & 0.3211 &  & 0.3252\\
3 & \underline{-0.9411\pm 0.0035} & -0.9552 &  & -0.9425\\
4 & 3.5034\pm 0.008 & \underline{3.4948} & 3.4269 & 3.4970\\
5 & -15.6620\pm 0.0049 & \underline{-15.6165\pm 0.0034} & -16.2018 & -15.6208
\end{array}
\]}
The relative precision on the results in the complex plane for the
chosen roots for respective orders \( p=2 \) to \( 5 \) are \(
9.\, 10^{-3},\, 4.\, 10^{-3},\, 6.\, 10^{-4}\, \textrm{and}\, 3.\,
10^{-4} \)
\\
\\
As far as I know, the criterion governing the choice for the best root still
works systematically for higher orders of perturbations.

\subsubsection*{IV.2.2. The vacuum energy in the strong coupling
limit}

The vacuum energy of the massless theory identifies to its strong
coupling limit: \( \lambda ^{1/3}\widetilde{E}_{0} \) where \(
\widetilde{E}_{0} \) is approximated by \(
\frac{2}{3}\overline{B}_{1} \). One can verify that the choice for
the best root made earlier still holds. The following array shows
the estimated values and their relative precisions compared to a
known value \( [11]\) of \( \widetilde{E}_{0} \) for order \( p=1
\) to \( 5 \):
\[
\begin{array}{lll}
p & \frac{2}{3}\overline{B}_{1} & \\ 1 & 0.420139 & (2.\,
10^{-3})\\ 2 & 0.4205216 & (1.\, 10^{-3})\\ 3 & 0.4208109\pm
0.0000952 & (2.\, 10^{-4})\\ 4 & 0.4207976 & (2.\, 10^{-5})\\ 5 &
0.4208087\pm 0.0000033i & (1.\, 10^{-5})\\ \,  &  & \\
\widetilde{E}_{0} & 0.4208049 &(\mbox{exact})
\end{array}\]
The accuracy obtained for the first order (where only the two
first terms of \( B \) are involved) is remarkably high (\( 0.2\%
\)) and increases with the order of perturbation. We could also
check that the choices for the best roots systematically hold for
all the coefficients \( \overline{B}_{n} \), and also for the sum
itself.

\subsubsection*{IV.2.3. The summation for finite coupling}

We also dispose of an analytical expression of the vacuum energy
in the case of a finite coupling (for any value of \( \lambda
=Z^{3} \) and \( m=1 \)). In order to compute numerically this
integral, the zeroth order term in powers of \( Z^{3} \) of the
integrant of the right sum (which we know to correspond to \( 1/2
\)) will have to be extracted from the integral as it is not
summable around \( t=0 \). As an example, we give the analytical
expression of the first order estimation for \(E(\lambda)\):
\[
E(\lambda)\sim\frac{1}{2}+2\int^{\infty}_0 \frac{dt}{t^2}\,
e^{-t^2} (e^{-\lambda(\varpi t)^3}-1)\frac{-1}{4\sqrt{\pi}}.
\]
where \(\varpi\simeq1.09954\). \\
 The following arrays show an
estimation of \( E(\lambda ) \) for \( \lambda =2 \), \( 4 \) and
\( 8 \) to be compared with the exact value \( [10,11] \). The
corresponding relative precisions are in brackets.
\[
\begin{array}{lll}
p & \lambda =2 & \\ 1 & 0.6957043 & (7.\, 10^{-4})\\ 2 & 0.6950365
& (2.\, 10^{-4})\\ 3 & 0.6951801\pm 0.0000363i & (6.\, 10^{-5})\\
4 & 0.6961732 & (3.\, 10^{-6})\\ 5 & 0.6961768\pm 0.0000008i &
(1.\, 10^{-6})\\ \,  &  & \\ E(2) & 0.6961758 &(\mbox{exact})
\end{array}\]
\\
\[
\begin{array}{lll}
p & \lambda =4 & \\ 1 & 0.8030005 & (1.\, 10^{-3})\\ 2 & 0.8035264
& (3.\, 10^{-4})\\ 3 & 0.8037776\pm 0.0000698i & (8.\, 10^{-5})\\
4 & 0.8037655 & (6.\, 10^{-5})\\ 5 & 0.8037727\pm 0.0000019i &
(4.\, 10^{-5})\\ \,  &  & \\ E(4) & 0.8037707 &(\mbox{exact})
\end{array}\]
\\
\[
\begin{array}{lll}
p & \lambda =8 & \\ 1 & 0.9504142 & (1.\, 10^{-3})\\ 2 & 0.9501856
& (4.\, 10^{-4})\\ 3 & 0.9505784\pm 0.0001160i & (1.\, 10^{-4})\\
4 & 0.9515597 & (1.\, 10^{-5})\\ 5 & 0.9515723\pm 0.0000034i &
(5.\, 10^{-6})\\ \,  &  & \\ E(8) & 0.9515685 &(\mbox{exact})
\end{array}\]
\\
As expected, the selected roots provide systematically the best
approximations, which reveal to improve greatly with the order of
perturbation. \\ One can also test successfully the procedure on
other observables of the model, such as the mean value of \( <\Phi
^{2}> \) or the excited states of energy. The same relative
precisions is achieved in these calculations.

Eventually, one can reasonably expect the method to work as well
for any \(x^{2N}\) potential in one dimension quantum mechanics.

\section*{V. Conclusion}

This paper does not provide any proof of the convergence of the
extension method despite the logic inherent to its construction.
Nevertheless, in the case of the anharmonic oscillator, it reveals
to have a highly predictive power for all positive values of the
coupling constant. It also makes a crucial and automatic link
between weak and strong couplings and provides an analytical
expression for the sum of some divergent perturbative series.
\\

Regarding the problems of divergences arising from quantum
mechanics and more generally from domains of physics facing
divergent expansions involving two physical parameters, the method
exposed here, besides its simplicity, has the advantage to be
systematic. It is also hoped to provide an excellent approximation
of a physical quantity (for any value of the coupling constant)
from the first perturbative correction. It opens perspectives for
field theories where only lower orders of perturbative corrections
(expected to be divergent) are available.\\

The power of the method obviously comes from our ability to take
into account physical information about dimensional quantities. If
we want to make it efficient for more complex models, or prove
that other classes of series suits the method, we will have to
extend it to models involving more than two physical parameters.
To do so, we will  manage to exhibit, in the same spirit as in
this work, constraints arising from dimensional information.  This
will be the subject of a second part of this work to be published.

\section*{Aknowledgement}

I want to thank Hugh F. Jones, François Molino and especially
Philippe Garcia for his long time support and contributions to
this work.

\pagebreak

\section*{Appendix: the extension procedure}

Given respective left an right forms of expansions:
\[
\sum ^{\infty}_{n=0}B_{-\alpha-\gamma n}\, Z^{-\alpha-\gamma
n}\,\,\,\,\textrm{and}\,\,\,\,\sum ^{\infty}_{n=0}B_{\delta n}\,
Z^{\delta n}\, ,
\]
We now summarize the method developed in this paper to estimate
the value to be associated to the known divergent truncated
series:
\[
\sum ^{p}_{n=0}B_{\delta n}\, Z^{\delta n}\, .
\]
\\
\textbf{The Signature:}
\\
We are in the general case of a \( \gamma |\alpha |\delta
\)-signature, the associated transformations is:
\[
 T_{n}=\frac{(-1-\frac{n}{\delta })!(-1+\frac{\alpha+n}{\gamma })!}{\Omega }\, .
\]
\\
\textbf{The Expansion:}
\\
\\
\( \begin{array}{l}
A_{\delta n}=T_{\delta n}^{-1}\, B_{\delta n}\, \, \, \, \, \,
0\leq n\leq p\\ \, \\ \, \, \, \, \, \, \, \, \, \, A^{i}(\omega
)=\sum ^{i}_{n=0}\frac{i!}{n!(i-n)!}(-1)^{n-i}\frac{A_{\delta
n}}{\omega ^{\delta n}}\, \, \, \, \, \, 0\leq i\leq p\\ \, \\ \,
\, \, \, \, \, \, \, \, \, \, \, \, \, \, \, \, \, \, \,
A^{p}(\omega )=0\, \, \, \Rightarrow \, \, \, \omega =\varpi \\ \,
\\ \, \, \, \, \, \, \, \, \, \, \overline{A}_{n}=\omega ^{n}\,
\sum ^{p-1}_{i=0}\frac{n!}{i!(n-i)!}A^{\ddot{i}}(\varpi )\, \, \,
\, \, \, n\in \hbox {\bm \char'132 }\\ \, \\
\overline{B}_{n}=T_{n}\, \overline{A}_{n}\, \, \, \, \, \, n\in
Z\\
\end{array} \)
\\
\\
\textbf{The Sum:}
\\
\[
\sum ^{\infty }_{n=0}A_{-\alpha - \gamma n }\,Z^{-\alpha - \gamma
n }\, \, \, \sim \, \, \, \gamma \, \int ^{\infty
}_{0}\frac{dt}{t}\,e^{-(t/\varpi Z)^{\gamma }-t^{\delta
}}(t/\varpi Z)^{\alpha }\, \sum ^{p-1}_{i=0}A^{i}(\varpi
)\frac{1}{i!}(-t^{\delta })^{i}\, \, \, .
\]
\\
which can also be written:
\[
 \sum ^{\infty }_{n=0}A_{\delta n\, }Z^{\delta n}\, \, \, \sim \,
\, \, \gamma \, \int ^{\infty }_{0}\frac{dt}{t}\,e^{-t^{\gamma
}-(\varpi Z t)^{\delta }}t^{\alpha }\, \sum
^{p-1}_{i=0}A^{i}(\varpi )\frac{1}{i!}(-(\varpi Z t)^{\delta
})^{i}\, \, \, .
\]
\\
\textbf{The Weight:}
\\
Between \( \{\omega _{1},\omega _{2},...,\omega _{p}\} \) roots of \( A^{p}(\omega ) \),
\( \varpi  \) is the one with the biggest weight:
\[
weight(\omega _{i})=\sum _{j\neq i}\frac{1}{(\omega _{j}-\omega _{i})^{2}}
\]
\pagebreak

\section*{References}

 \( ^{1} \)J. J. Loeffel, A. Martin, B. Simon, A. S. Wightman,
Phys. lett. \textbf{B 30}, 656 (1969).
\\
\( ^{2} \)A. Neveu, Nucl. Phys. B, \textbf{18B}, 242 (1990). \\
\\
\( ^{3} \)B. Bellet, P. Garcia, A. Neveu, Int. J. Mod. Phys.
\textbf{A 11}, 5587 (1996).\\
\\
\( ^{4} \)R. Seznek, J. Zinn-Justin, J. Math. Phys.
\textbf{20}(7), 1398 (1979).\\
\\
\( ^{5} \)I. R. C. Buckley, A. Duncan, H. F. Jones Phys. Rev.
\textbf{D47}, 2554 (1993).\\
\\
\( ^{6} \)R. Guida, K. Konishi, H. Suzuki, Ann. Phys.
\textbf{241}, 152 (1995).\\
\\
\( ^{7} \)P. M. Stevenson Phys. Rev. \textbf{D 23}, 2916 (1981).\\
\\
\( ^{8} \)C. M. Bender, L. M. A. Bettencourt, Phys. Rev. \textbf{D
54}, 7710 (1996).\\
\\
\( ^{9} \)B. Simon, Ann. Phys.  \textbf{58}, 76 (1970).\\
\\
\( ^{10} \)W. Janke, H. Kleinert, Phys. Lett. \textbf{75}, 2787
(1995).\\
\\
\( ^{11} \)F. Vinette, J. Cizek, J. Math. Phys. \textbf{32}, 3392,
(1991).\\
\\
\( ^{12} \)C. M. Bender, T. T. Wu, Phys. Rev. \textbf{184}, 1231
(1969).

\end{document}